\newcommand\FAP{5FAP-CuCl$_4$}
\documentclass[twocolumn,prb,showpacs,preprintnumbers,amsmath,amssymb,superscriptaddress]{revtex4}

\usepackage{graphicx}
\usepackage{dcolumn}
\usepackage{bm}
\usepackage{epsfig}
\usepackage{color}

\begin{document}

\title{Dynamics of the two-dimensional $S=1/2$ dimer system $\rm (C_5H_6N_2F)_2CuCl_4$}

\author{Tao Hong}
\affiliation{Neutron Scattering Sciences Division, Oak Ridge National Laboratory, Oak Ridge, Tennessee 37831-6393, USA}
\author{S.~N.~Gvasaliya}
\affiliation{Laboratory for Neutron Scattering, Paul Scherrer
Institute, Villigen PSI, CH-5232, Switzerland}
\author{S.~Herringer}
\affiliation{Carlson School of Chemistry and Department of Physics, Clark University, Worcester, Massachusetts 01610, USA}
\author{M. M. Turnbull}
\affiliation{Carlson School of Chemistry and Department of Physics, Clark University, Worcester, Massachusetts 01610, USA}
\author{C. P. Landee}
\affiliation{Carlson School of Chemistry and Department of Physics, Clark University, Worcester, Massachusetts 01610, USA}
\author{L.-P. Regnault}
\affiliation{CEA-Grenoble, INAC-SPSMS-MDN, 17 rue des Martyrs, 38054 Grenoble Cedex 9, France}
\author{M.~Boehm}
\affiliation{Institut Laue-Langevin, B.P. 156, F-38042 Grenoble, France}
\author{A.~Zheludev}
\affiliation{Neutron Scattering and Magnetism, Laboratorium
f$\rm\ddot{u}$r Festk$\rm\ddot{o}$rperphysik, ETH
Z$\rm\ddot{u}$rich, Z$\rm\ddot{u}$rich CH-8093, Switzerland}

\date{\today}

\begin{abstract}
Inelastic neutron scattering was used to study a quantum
\emph{S}=1/2 antiferromagnetic Heisenberg
system---Bis(2-amino-5-fluoropyridinium) Tetrachlorocuprate(II). The
magnetic excitation spectrum was shown to be dominated by long-lived
excitations with an energy gap of $\Delta$=1.07(3) meV. The measured
dispersion relation is consistent with a simple two-dimensional
square lattice of weakly-coupled spin dimers. Comparing the data to
a random phase approximation treatment of this model gives the
intra-dimer and inter-dimer exchange constants $J$=1.45(2) meV and
$J^\prime$=0.31(3) meV, respectively.
\end{abstract}


\maketitle

Quantum spin liquids, where strong zero point fluctuation destroy
magnetic long-range order even at $T$=0, have been a subject of
intense studies for over two decades. Of particular interest are two-dimensional systems of this type. In addition to their relevance to layered superconducting cuprates,\cite{Anderson87:235,Lee06:78} they are of interest in the context of field-induced quantum phase transitions,\cite{Sach99,Thier08:4} where, to date, most
experimental work was performed on either spin chains\cite{Stone03:91,Hong09:80} and ladders,\cite{Masuda06:96,Hong10:105} or 3D systems.\cite{Stone02:65,Ruegg03:423} Several 2D spin liquids including $\rm BaCuSi_2O_6$,\cite{Sebastian06:441} $\rm SrCu_2(BO_3)_2$,\cite{Gaulin04:93} and PHCC\cite{Stone01:64} have been studied
in great detail, but show rather complex behavior due to geometrically frustrated interactions.

\begin{figure}
\includegraphics[width=7.5cm,bbllx=50,bblly=80,bburx=555,
  bbury=760,angle=0,clip=]{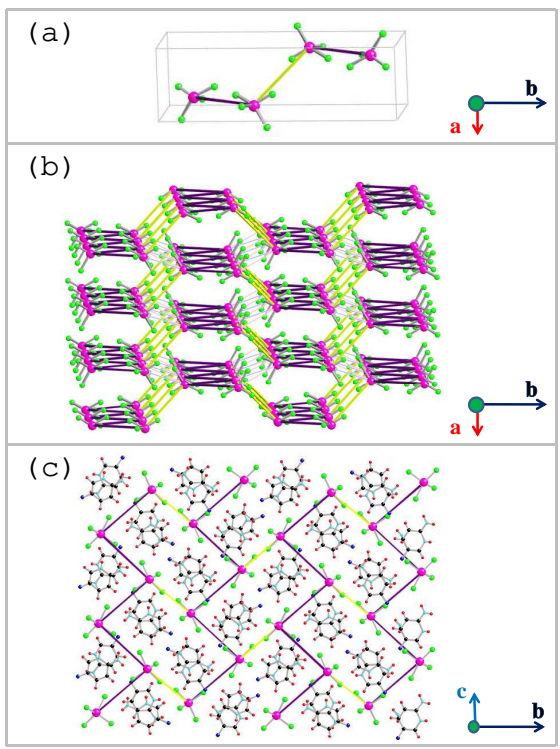}
\caption{(color online) (a) Outline of a unit cell in~\FAP. (b) View
from the \textbf{\emph{c}} axis, showing the zigzag layer structure.
Only Cu and Cl atoms are shown, the organic cations have been
removed for clarity. (c) View from the \textbf{\emph{a}} axis,
showing the arrangement of $\rm CuCl_4^{2-}$ anions and organic
cations. Different color lines stand for the different magnetic
interactions. Color coding is as follows: \textcolor{magenta}{Cu},
\textcolor{green}{Cl}, \textcolor{black}{C}, \textcolor{red}{H},
\textcolor{blue}{F}, and \textcolor{cyan}{N}.} \label{fig1}
\end{figure}

Recently, a new quantum antiferromagnet Bis(2-amino-5-fluoropyridinium) Tetrachlorocuprate(II) (\FAP, for short) has been discovered and characterized by bulk measurements.\cite{Li07:46} The ground state is a spin liquid. The
energy gap $\Delta$ to the lowest excited (triplet) state was determined to be 0.97(4) meV from the high-field magnetization data. \FAP~($\rm (C_5H_6N_2F)_2CuCl_4$) crystallizes in a monoclinic space group \emph{P}2$_1$/c with lattice constants $a$=6.926(7) $\AA$, $b$=21.73(2) $\AA$, $c$=10.911(10) $\AA$, and $\beta$=100.19(2)$^\circ$. There are four inequivalent Cu atoms per unit cell as shown in Fig.~\ref{fig1}(a). In view of crystal structure,~\FAP~consists of $\rm CuCl_4^{2-}$ anion layers separated by organic cations. Those $\rm CuCl_4^{2-}$ anions form zigzag
layers which are stacked along the the crystallographic \textbf{\emph{a}} axis as shown in Fig.~\ref{fig1}(b).
Fig.~\ref{fig1}(c) shows the arrangement of $\rm CuCl_4^{2-}$ anions and organic cations viewing from the \textbf{\emph{a}} axis. The yellow and purple lines indicate the most plausible super-exchange
interactions that could be carried across Cu-Cl-Cl-Cu bonds. The former involves the shortest Cl-Cl contact with a separation of 3.66 $\AA$, while for the latter the relevant Cl-Cl distance is 4.07 $\AA$. One can expect inter-layer coupling along the \textbf{\emph{a}} axis to be exponentially weak due to a much larger Cl-Cl separation 5.07 $\AA$. Even with this in mind, it is not obvious which in-lane interactions are the most relevant. To better understand the spin Hamiltonian for~\FAP, in this paper, we report inelastic neutron scattering (INS) measurements of magnetic
excitations and their dispersion relation. We show that~\FAP~can be
adequately described by a minimal model that is equivalent to a
square lattice of weakly interacting antiferromagnetic $S=1/2$
dimers with no geometric frustration.

Small size fully deuterated single crystalline samples of~\FAP~were
prepared by the procedure described in Ref.~\onlinecite{Li07:46}.
About twenty single crystals ($\sim$0.3 g total) were co-aligned
with a mosaic about 4$^\circ$ and used for INS measurements on the
cold-neutron triple-axis spectrometer IN14 at Institut
Laue-Langevin. Another assembly consisting of 20 single crystals
($\sim$0.5 g) with a total 5$^\circ$ mosaic was used for the
measurements on TASP\cite{Semadeni01:297} spectrometer at SINQ, Paul
Scherrer Institute. Horizontal beam divergences were given by
$^{58}$Ni guide-60'-open-open at IN14 and by $10'/ \AA$-open-open-open at TASP.
Both experiments were performed with the
scattered wave vector \textbf{$k_f$} fixed at 1.5 $\AA^{-1}$ and Be
filter placed after the sample was used to remove higher order beam
contamination. The sample was oriented horizontally in a
(0,\emph{k},\emph{l}) reciprocal lattice plane and cooled in a
standard He-4 cryostat.

\begin{figure}
\includegraphics[width=7.5cm,bbllx=95,bblly=55,bburx=545,bbury=730,angle=0,clip=]{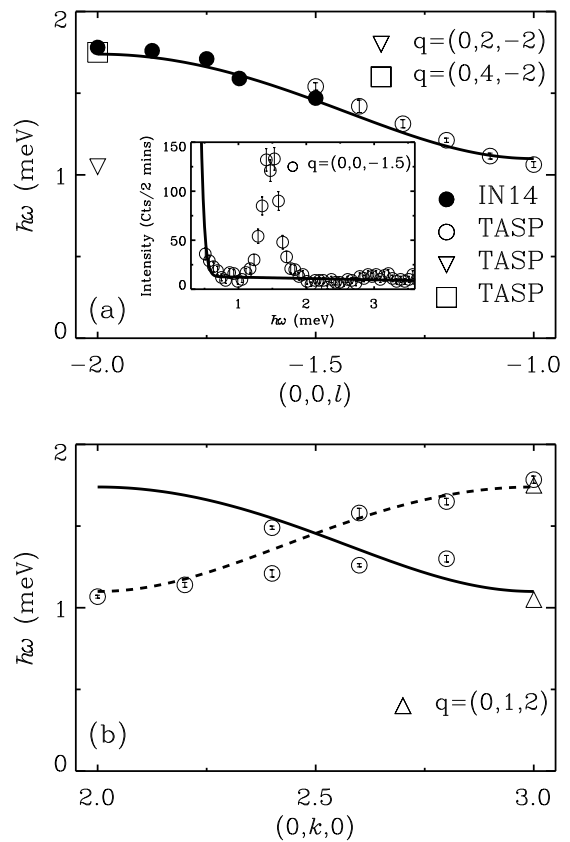}
\caption{Dispersion of magnetic excitations measured in~\FAP~at
$T=1.5$~K as a function of $k$ and $l$. The solid (dashed) line is a
fit to the acoustic (optical) magnon dispersion mode as discussed in
the text. Inset: Constant-$\mathbf{q}=(0, 0, -1.5)$ scan as a
function of the transferred energy $\hbar\omega$ at $T =1.5$~K. The
solid line is a Gaussian and a linear term fit to account for the
background contribution.}
\label{fig2}
\end{figure}

\begin{figure}
\includegraphics[width=7.5cm,bbllx=35,bblly=90,bburx=500,bbury=620,angle=0,clip=]{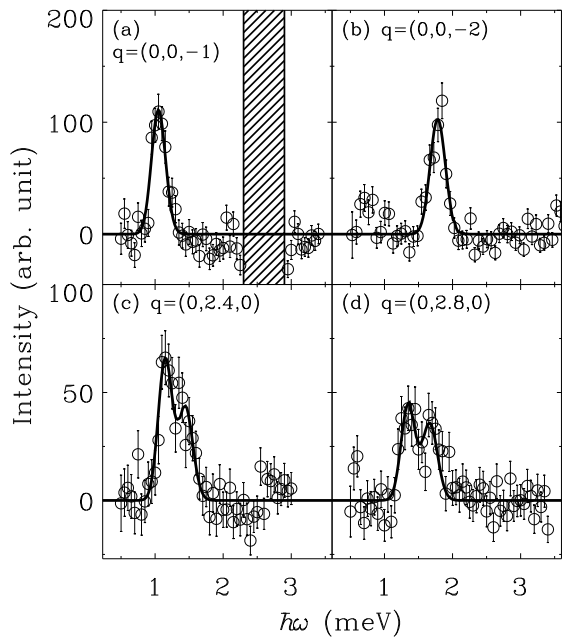}
\caption{Representative background-subtracted INS data for~\FAP~at
$T=1.5$~K. The solid lines are fits Gaussian profiles of
instrumental width. The shaded area in (a) is excluded due to a
contamination by ``accidental Bragg'' spurious scattering.} \label{fig3}
\end{figure}

The data were collected in a series of constant-\textbf{\emph{q}}
scans at $T=1.5$ K along the high symmetry directions in the
(0,\emph{k},\emph{l}) reciprocal plane. The inset of
Fig.~\ref{fig2}(a) shows the raw data of a typical scan performed at
\textbf{\emph{q}}=(0,0,-1.5), where a sharp resolution-limited peak
due to a magnetic excitation is clearly visible. The background
(shown as a solid line)  was determined by fitting the data over the
range where no magnetic excitations are apparent, to a term linear
in energy, plus a resolution-limited Gaussian peak at zero energy
transfer. The latter accounted for incoherent elastic scattering.
Background-subtracted data at several wave vectors are shown in
Fig.~\ref{fig3}. At certain wave vectors the observed signal has a
distinct double-peak structure as shown in Figs.~\ref{fig3}(c) and
(d). Solid lines in Fig.~\ref{fig3} are fits to Gaussian profiles of
fixed widths, determined by instrumental resolution, as calculated
using the Cooper-Nathans approximation.\cite{Cooper67:23} Excitation energies (dispersion relation)
obtained from such fits are summarized in Fig.~\ref{fig2}. The
spectrum appears to consist of two equivalent branches, an ``optic''
and an ``acoustic'' one, although the former is not visible at the
majority of the wave vectors studied.

\begin{figure}
\includegraphics[width=6.5cm,bbllx=45,bblly=65,bburx=525,bbury=700,angle=-90,clip=]{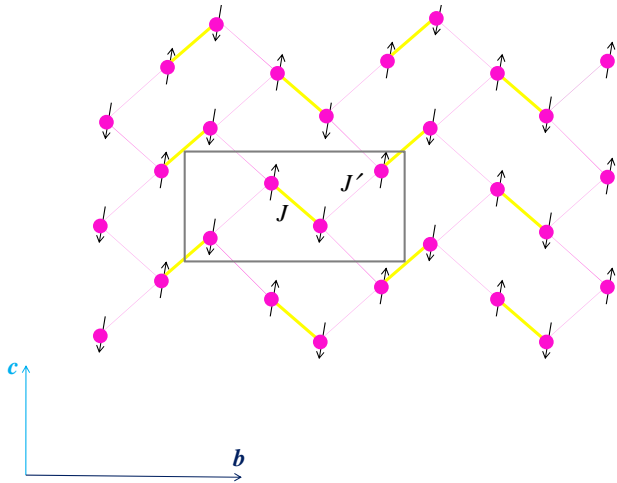}
\caption{(color online) Minimal model for magnetic interactions in~\FAP, shown in
projection onto the $(b,c)$ plane. Dimer singlets are formed by
antiferromagnetic exchange $J$. Dispersion of singlet-triplet
excitations is due to inter-dimer interaction $J^\prime$.}
\label{fig4}
\end{figure}

The most comprehensive method of determining the relevant exchange
parameters in a gapped system with weakly dispersive excitations is
based on the application of the first-moment sum rule for the
magnetic dynamic structure factor.\cite{Hohenberg74:10} This
procedure involves quantitative measurements of inelastic
intensities as a function of wave vector and was successfully
applied to the rather complex interaction geometry in
CuHpCl.\cite{Stone02:65} Due to technical complications,
associated with the sample fracturing upon cooling, and the data
being collected in several disjoint experiments, in the present
study we adopted a simpler approach. Only the measured excitation
energies were analyzed. A parameterized model was refined to
reproduce the experimental dispersion relations. The proposed
minimal model is visualized in Fig.~\ref{fig4}, which shows the
exchange constants a single layer of magnetic Cu$^{2+}$ ions. The
dimers are formed by the stronger antiferromagnetic bonds $J$, which
ensures a spin-singlet ground state and an energy gap. The
dispersion of singlet-triplet dimer excitations is due to  the
weaker inter-dimer exchange constant $J^\prime$. Topologically, this
spin network can be viewed as a square lattice of $S=1/2$ dimers.
Nevertheless, its actual realization in the crystal has two dimers
per crystallographic unit cell. This leads to a folding of the
Brillouin zone and the appearance of two excitation branches. As
shown below, this minimal model seems to be fully consistent with
experimental observations.

We can treat the weak inter-dimer coupling $J^\prime$ at the Random
Phase Approximation (RPA) level.\cite{Haley72:5} This method has
been applied to numerous spin systems, including $\rm
Cs_3Cr_2Br_9$,\cite{Leue84:30} $\rm BaCuSi_2O_6$,\cite{Sasago97:55}
$\rm BaCuSi_2O_7$,\cite{Zhel00:62} $\rm
Ba_3Mn_2O_8$,\cite{Stone08:77} and $\rm
Ba_3Cr_2O_8$.\cite{Kofu09:102-1} In application to~\FAP~, we can
treat two decoupled dimers in a unit cell as the basic spin cluster
(tetramer). These tetramers will then form a Bravais lattice. With
four sites in each tetramer, their bare susceptibilities are written
as a $4\times 4$ matrix, the indexes enumerating the four spins in
each unit:
\begin{align}\label{chitetra}
\chi_{0}(\omega)=\frac{J}{2J^2-2(\hbar\omega)^2} R(T)
\begin{pmatrix}
1 & -1 & 0 & 0 \\
-1 & 1 & 0 & 0 \\
0 & 0 & 1 & -1 \\
0 & 0 & -1 & 1 \\
\end{pmatrix}.
\end{align}
Here $R(T)=\frac{1-e^{-J/(k_BT)}}{1+3e^{-J/(k_BT)}}$ is the
difference in population of the excited and ground states of the dimers. The weak interaction $J'$ couples adjacent tetramers, but also dimers within each tetramer. Its effect is taken into account within the RPA. The corresponding
RPA equation is written as:
\begin{eqnarray}\label{rpa}
\chi^{RPA}(\textbf{q},\omega)=\chi_{0}(\omega)
 [1-\chi_{0}(\omega)J^\prime(\textbf{q})]^{-1}.
\end{eqnarray}
Here $J^\prime(\textbf{q})$ is the Fourier transform of exchange
interactions due to $J'$. It is also written as a $4\times 4$
matrix:
\begin{align}\label{jq}
J^\prime(\textbf{q})=J^\prime \Biggl(\begin{smallmatrix}
0 & 0 & 0 & e^{-i2\pi (k+l)}+e^{-i2\pi k} \\
0 & 0 & 1+e^{-i2\pi l} & 0 \\
0 & 1+e^{i2\pi l} & 0 & 0 \\
e^{i2\pi (k+l)}+e^{i2\pi k} & 0 & 0 & 0 \\
\end{smallmatrix}\Biggr).
\end{align}

The dispersion relation is then readily obtained from the location
of poles in $\chi^{RPA}(\textbf{q},\omega)$:
\begin{align}\label{rpadis}
(\hbar\omega_{\mathbf{q}})^2=J^2\mp JJ^\prime
R(T)\left[\cos\pi(k+l)+\cos\pi(k-l)\right].
\end{align}

The solid and dashed lines in Fig.~\ref{fig2} are fits of this
dispersion relation to the experimental data. A good agreement is
obtained with $J=1.45(2)$ meV and $J^\prime=0.31(3)$ meV.

In summary, \FAP~can be described as a simple 2D network of weakly
interacting spin dimers, which is topologically equivalent to a
dimer square lattice. The magnon dispersion bandwidth is 0.70(4)~meV,
smaller than gap energy $\Delta=1.07(3)$ meV, but still substantial.
Future work will focus on the high-field study of quantum phase
transition and critical phenomena in~\FAP.

\begin{acknowledgments}
One of the authors (AZ) would like to thank B. Normand for
discussions. The work at ORNL was partially funded by the Division of Scientific User Facilities, Office of Basic Energy Sciences - Materials Science, Department of Energy.
\end{acknowledgments}

\end{document}